\documentclass[letterpaper,10pt]{article}
\usepackage{setspace}
\usepackage{graphicx}
\usepackage{amsmath}
\usepackage{color}
\usepackage{caption}
\usepackage{array}
\usepackage[sectionbib,square]{natbib}
\usepackage{verbatim} 	
\usepackage{amssymb}

\usepackage[letterpaper,vmargin=1in,hmargin=1in]{geometry}

\newcommand{\fignameSP}{Fig.~}
\newcommand{\fignamesSP}{Figs.~}
\newcommand{\subfig}{-}
\newcommand{\tabnameSP}{Table }

\newcommand{\ALPM}{ALPM}
\newcommand{\DEM}{DEM}
\newcommand{\EBE}{EBE}
\newcommand{\RC}{RC}
\newcommand{\SDE}{SDE}
\newcommand{\EBEArea}{A^e}
\newcommand{\AreaReinf}{A_s}
\newcommand{\BaseBeam}{b_b}
\newcommand{\BaseCol}{b_c}
\newcommand{\Bend}{B}
\newcommand{\YieldBend}{B_y}
\newcommand{\DiamBeam}{D_b}
\newcommand{\DiamCol}{D_c}
\newcommand{\YoungERC}{E_c}
\newcommand{\YoungESteel}{E_s}
\newcommand{\CStrengthEC}{f_c}
\newcommand{\YieldStressSteel}{f_y}
\newcommand{\Gravity}{g}
\newcommand{\HeightCross}{h}
\newcommand{\HeightBeam}{h_b}
\newcommand{\HeightCol}{h_c}
\newcommand{\Height}{H}
\newcommand{\TotHeight}{H_{tot}}
\newcommand{\EBEInertia}{I^e}
\newcommand{\Bay}{L}
\newcommand{\TotWidth}{L_{tot}}
\newcommand{\EBELen}{L^e}
\newcommand{\NCell}{n}
\newcommand{\AxForce}{N}
\newcommand{\YieldNT}{N_y}
\newcommand{\YieldNC}{N_{cy}}
\newcommand{\ExtLoad}{q_{ext}}
\newcommand{\UltLoadInt}{q_u^I}
\newcommand{\CritLoad}{q_c}
\newcommand{\PartTotLoad}{q_{p,t}}
\newcommand{\EqLoad}{q_{eq}}
\newcommand{\ROne}{R_1}
\newcommand{\TensReinfFrac}{t_s}
\newcommand{\Shear}{T}
\newcommand{\SpecWeiRC}{\gamma_{RC}}
\newcommand{\EBEDampL}{\gamma_L}
\newcommand{\EBEDampB}{\gamma_B}

\newcommand{\CrossBeamAspRatio}{\delta_b}
\newcommand{\ExtraYieldBend}{\Delta B_y}
\newcommand{\AxStrain}{\varepsilon}
\newcommand{\StrainReinf}{\varepsilon_s}
\newcommand{\UltStrainRC}{\varepsilon_{u,c}}
\newcommand{\UltStrainSteel}{\varepsilon_{u,s}}
\newcommand{\PlastAxStrain}{\varepsilon^{pl}}
\newcommand{\EBEYieldEpsT}{\varepsilon_y}
\newcommand{\EBEYieldEpsC}{\varepsilon_{cy}}
\newcommand{\EloThresh}{\varepsilon_{th}}
\newcommand{\CompThresh}{\varepsilon_{cth}}
\newcommand{\SlendCell}{\lambda}
\newcommand{\SlendBeam}{\lambda_b}
\newcommand{\SlendCol}{\lambda_c}

\newcommand{\ReinfFrac}{\rho_s}
\newcommand{\ReinfFracBeam}{\rho_{s,b}}
\newcommand{\ReinfFracCol}{\rho_{s,c}}
\newcommand{\DiamBar}{\phi}
\newcommand{\Rot}{\varphi}
\newcommand{\RotPlast}{\varphi^{pl}}
\newcommand{\EffRot}{\varphi^{eff}}
\newcommand{\YieldEffRot}{\varphi^{eff}_y}
\newcommand{\RotThresh}{\varphi_{th}}

\begin{document}

\title{Hierarchical structures for a robustness-oriented capacity design}
\maketitle

\begin{center}
\author{E. Masoero\footnotemark[1], 
F. K. Wittel\footnotemark[2], 
H. J. Herrmann\footnotemark[3], 
B. M. Chiaia\footnotemark[4].
}
\end{center}

\footnotetext[1]{Dr., Politecnico di Torino, Department of Structural and Geotechnical Engineering, Corso Duca degli Abruzzi 24, 10129 Torino, Italy. Massachusetts Institute of Technology, Department of Civil and Environmental Engineering, 77 Massachusetts Avenue, 02139, Cambridge, MA, U.S.A. Email address: emasoero@mit.edu}
\footnotetext[2]{Dr., ETH Zurich, Institute for Building Materials, Schafmattstrasse 6, 8093 Zurich, Switzerland. Email address: fwittel@ethz.ch}
\footnotetext[3]{Prof., ETH Zurich, Institute for Building Materials, Schafmattstrasse 6, 8093 Zurich, Switzerland. Email address: hans@ifb.baug.ethz.ch}
\footnotetext[4]{Prof., Politecnico di Torino, Department of Structural and Geotechnical Engineering, Corso Duca degli Abruzzi 24, 10129 Torino, Italy. Email address: bernardino.chiaia@polito.it}
\begin{abstract}
In this paper, we study the response of 2D framed structures made of rectangular cells, to the sudden removal of columns. We employ a simulation algorithm based on the Discrete Element Method, where the structural elements are represented by elasto-plastic Euler Bernoulli beams with elongation-rotation failure threshold. The effect of structural cell slenderness and of topological hierarchy on the dynamic residual strength after damage $\ROne$ is investigated. Topologically \textit{hierarchical} frames have a primary structure made of few massive elements, while \textit{homogeneous} frames are made of many thin elements. We also show how $\ROne$ depends on the activated collapse mechanisms, which are determined by the mechanical hierarchy between beams and columns, i.e. by their relative strength and stiffness. Finally, principles of robustness-oriented capacity design which seem to be in contrast to the conventional anti-seismic capacity design are addressed.
\end{abstract}


\section*{Introduction}\label{Intro}
Since many decades, design codes ensure a very low probability that a building collapses under ordinary loads, like self weight, dead and live service load, or snow. Nevertheless buildings still do collapse, from time to time. An extremely small fraction of collapses originates from unlikely combinations of intense ordinary load with very poor strength of the building. The majority of structural collapses are due to accidental events that are not considered in standard design. Examples of such events are: gross design or construction errors, irresponsible disregard of rules or design prescriptions, and several rare load scenarios like e.g.~earthquakes, fire, floods, settlements, impacts, or explosions \citep{Alexander-2004}. Accidental events have low probability of occurrence, but high potential negative consequences. Since risk is a combination of probability and consequences, the risk related to accidental events is generally significant \citep{Eurocode_1-1-7}. 

In 1968 a gas explosion provoked the partial collapse of the Ronan Point building in London. This event highlighted for the first time the urgency for \textit{robust} structures, enduring safety in extraordinary scenarios \citep{Pearson_Delatte-2005}. Since then, interes was driven by striking catastrophic collapses \citep{Val_Val-2006}, until in 2001 the tragic collapse of the World Trade Center renewed the attention to the topic (see e.g.~\citep{Bazant_Zhou-2002} and \citep{Cherepanov_Esparragoza-2007}). The last decades, several design rules aimed at improving structural robustness have been developed (see e.g.~\citep{Masoero-PHD-2010}).

Accidental events can be classified into \textit{identified} and \textit{unidentified} \citep{Eurocode_1-1-7,DoD_UFC-2005}. Identified events are statistically characterizable in terms of intensity and frequency of occurrence. Examples are earthquakes, fire not fueled by external sources, gas explosions, and unintentional impacts by ordinary vehicles, airplanes, trains, or boats. Specific design rules and even entire codes are devoted to specific identified accidental events. Unidentified events comprise a wide variety of incidents whose intensity and frequency of occurrence can not be described statistically, e.g.~terrorist attacks or gross errors. 

The risk related to unidentified accidental events can be mitigated both by structural and nonstructural measures \citep{Gulvanessian-2006}. Nonstructural measures such as barriers and monitoring can reduce the probability that an accidental event affects the structural integrity, others like a wise distribution of plants and facilities can minimize the negative consequences of eventual collapses. Otherwise, structural measures can improve local resistance of structural elements to direct damage, e.g.~the design of \textit{key elements} for intense local load \citep{Eurocode_1-1-7}, or the application of the \textit{Enhanced Local Resistance} method \citep{DoD_UFC-2005}. Structural measures can also provide progressive collapse resistance, i.e.~prevent spreading of local direct damage inside the structure to an extent that is disproportioned with respect to the initial event. Usual strategies to improve progressive collapse resistance are compartmentalization of structures \citep{Starossek-2006} and delocalization of stress after local damage. Stress delocalization can be obtained exploiting redundancy, plastic stress redistributions (Masoero, Wittel et al., 2010), ties \citep{Alexander-2004}, and moment resisting connections \citep{Hamburger_Whittaker-2004,Vlassis_Izzudin-2008}.

Nowadays several design codes employ the conventinal \textit{Alternate Load Path Method (\ALPM)} to evaluate progressive collapse resistance, e.g.~\citep{GSA-2003} and \citep{DoD_UFC-2005}. The method consists in removing one key element, generally a column or a wall, and measuring the extent of subsequent collapse. If the final collapse is unacceptably wide, some of the previously listed measures have to be employed. Hence structures are first designed and subsequently tested to be robust - they are not conceived \textit{a priori}. This course of action excludes optimizations of the basic structural topology and geometry, that actually play a key role in the response to local damage, considering as an example the very different behavior of redundant and statically determined structures. Anti-seismic design \citep{Eurocode_8} already contains some prescriptions that should be considered before starting a new design, e.g.~geometric regularity on the horizontal and on the vertical planes. Furthermore, anti-seismic \textit{capacity design} requires a hierarchy of the structural elements ensuring that earthquakes can only provoke ductile collapse of the horizontal beams, while failure of columns and brittle ruptures due to shear are inhibited. Differently, for what concerns progressive collapse resistance, optimal overall geometric features are not known, except for the concepts of redundancy and compartmentalization. Furthermore, the idea of hierarchically maximizing progressive collapse resistance is completely absent. 

In this paper, we make a first step to cover this deficiency, showing that progressive collapse resistance can be improved by hierarchy in the overall geometry (\textit{topological} hierarchy) and in the relative strength and stiffness of horizontal and vertical structural elements (\textit{mechanical} hierarchy). Our approach incorporates the simulation of progressive collapse of regular 2D frames made of reinforced concrete (\RC) subjected to the sudden removal of structural elements, following the \ALPM~framework. We first describe the analyzed frame structures and briefly sketch the approach that is based on the \textit{Discrete Element Method (\DEM)}. After the model description, we present the results of the simulations, with focus on the effect of geometry and hierarchy on the activated collapse mechanisms and, consequently, on progressive collapse resistance.
\section*{Hierarchical structures and damage}
\begin{figure}[htb]
\begin{center}
\includegraphics{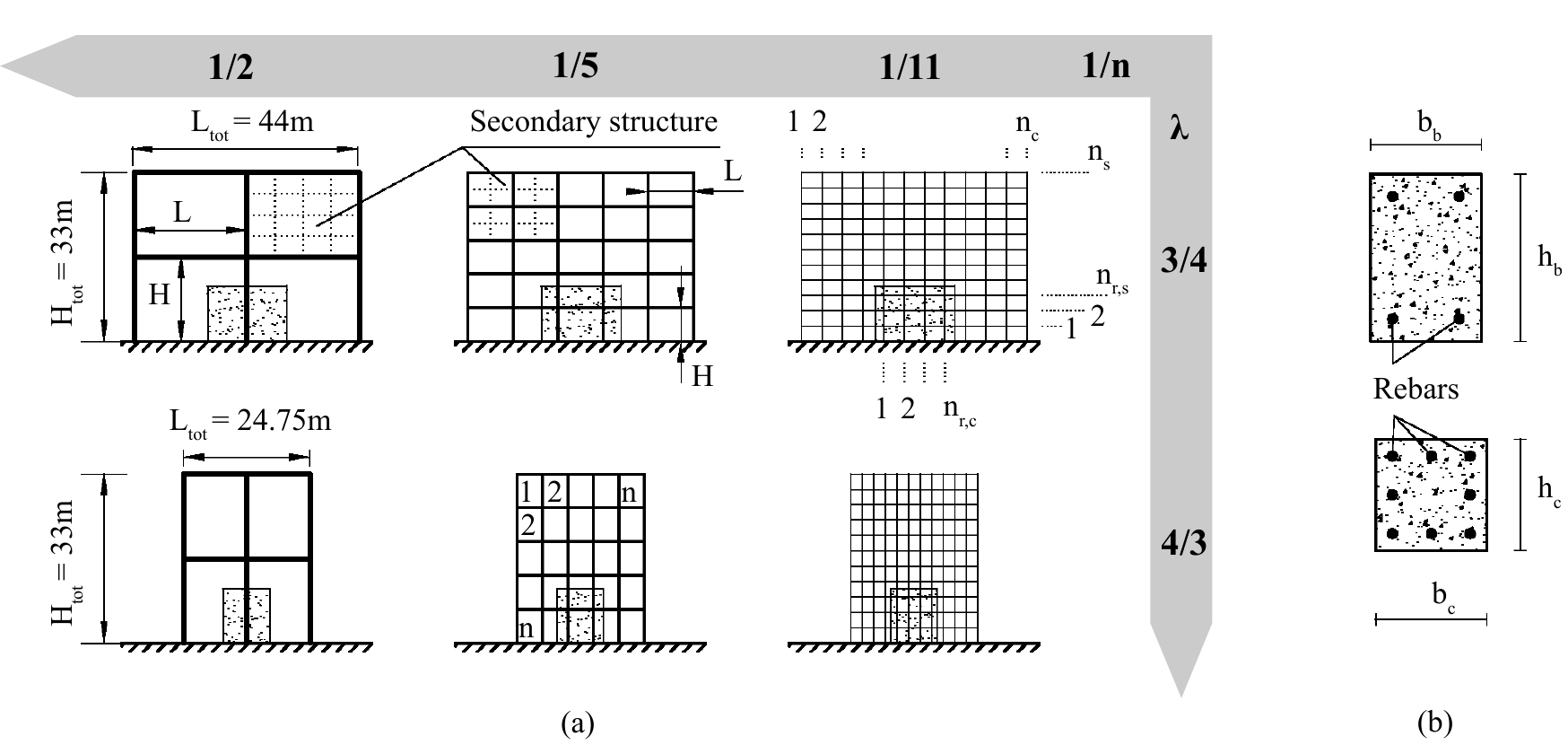}
\caption{a) Map of the studied frames. The cohesion of the elements inside the dotted damage area is suddenly removed to represent the initial accidental damage. b) Cross sections and rebars of beams (above) and columns (below).}\label{Fig_Struct}
\end{center}
\end{figure}

We consider two representative sets of regular 2D framed structures in \fignameSP\ref{Fig_Struct}\subfig a. Each set consists of three frames with identical total width $\TotWidth$ and different topological \textit{hierarchical level} $1/\NCell$, where $\NCell^2$ is the number of structural cells in a frame. The horizontal beams, excluded those of the secondary structure, carry a uniform load per unit length $\ExtLoad$. The frames are made of \RC~with typical mechanical parameters of concrete and steel, as shown in \tabnameSP\ref{tabMecPar}. The total height $\TotHeight$ of the structure is kept constant, and two different height-bay aspect ratios $\SlendCell=\Height/\Bay$ of the structural cells are considered ($\SlendCell=0.75$ and $\SlendCell=1.33$). 
\begin{table}[htb]
	\begin{center}
{
\begin{tabular}{|p{4.4cm}|>{\centering\arraybackslash}p{1.1cm}|>{\centering\arraybackslash}p{0.8cm}|>{\centering\arraybackslash}p{1.1cm}|}
\hline
$Parameter$   			&    $Symbol$    			&   $Units$ 		& $Value$ 	\\
\hline
\hline
\multicolumn{4}{|l|}{$Reinforced \; concrete$} \\
\hline
Specific weight			&   	$\SpecWeiRC$		&  kg/m$^3$		& 2500\\
Young modulus			&   	$\YoungERC$     		&  N/m$^2$   		& 30$\cdot 10^9$ \\
Compressive strength (high)	&   	$\CStrengthEC$		&  N/m$^2$		& 35$\cdot 10^6$ \\
Compressive low (low) 	& 	$\CStrengthEC$		&  N/m$^2$		& 0.35$\cdot 10^6$ \\
Ultimate shortening 		&   	$\UltStrainRC$		&  -				& 0.0035	\\
\hline
\multicolumn{4}{|l|}{$Steel$} \\
\hline
Young's modulus		&  	$\YoungESteel$		&  N/m$^2$		& 200$\cdot 10^9$ \\
Yield stress			&   	$\YieldStressSteel$	&  N/m$^2$		& 440$\cdot 10^6$\\
Ultimate strain			&   	$\UltStrainSteel$		&  -				& 0.05 \\
\hline
\end{tabular}
}	\caption{Mechanical properties of reinforced concrete and steel.}\label{tabMecPar}
	\end{center}
\end{table}

There exist several ways of introducing hierarchy into the topology of framed structures; here we call a structure ``hierarchical" if it has a primary structure, made of few massive structural elements, that supports a secondary one. The latter defines the living space and has negligible stiffness and strength compared to the primary structure. The frames with $\NCell=2$ and $\NCell=5$ can be seen as reorganizations of those with $\NCell=11$. In detail, each column of the frames with $\NCell=5$ corresponds to two columns of the frames with $\NCell=11$, and the same is valid for the beams, disregarding the first floor beam of the frames with $\NCell=11$, which is simply deleted (see \fignameSP\ref{Fig_Struct}\subfig a). Analogously, the geometry of the frames with $\NCell=2$ can be obtained starting from the frames with $\NCell=5$.

The cross sections of columns are square (see \fignameSP\ref{Fig_Struct}\subfig b), with edges $\HeightCol=\BaseCol$ proportional to $\Height$ with factor $\SlendCol=1/10$. The beams have rectangular cross section whose height $\HeightBeam$ is proportional to $\Bay$ with factor $\SlendBeam=1/10$, and whose base $\BaseBeam$ is proportional to $\HeightBeam$ with aspect ratio $\CrossBeamAspRatio=2/3$. The reinforcement is arranged as shown in \fignameSP\ref{Fig_Struct}\subfig b, with area $A_s$ proportional to the area of the cross section by factor $\ReinfFracCol=0.0226$ for the columns (i.e.~8$\DiamBar$18 when \NCell=11), and $\ReinfFracBeam=0.0029$ for the beams (i.e.~4$\DiamBar$14 when \NCell=11).

The damage areas, (dotted in \fignameSP\ref{Fig_Struct}\subfig a) contain the structural elements that are suddenly removed to represent an accidental damage event, following the \ALPM~framework. The damage is identical for frames with same $\SlendCell$, and is defined by the breakdown of one third of the columns on a horizontal line. The columns and beams removed from frames with $\NCell=11$ correspond to the structural elements removed from frames with $\NCell=5$ and $\NCell=2$. This kind of damage is employed to represent accidental events with a given amount of destructive energy or spatial extent, like explosions or impacts. In this work we do not explicitly simulate very local damage events like gross errors, which would be better represented by the removal of single elements. Nevertheless, we will generalize our results to consider also localized damage events.
%
%
%
%
%
\section*{DEM model}
We employ the Discrete Element Method (DEM) to simulate the dynamics after a sudden damage \citep{Poschel_Schwager-2005,Carmona_Wittel-2008}. DEM is based on a Lagrangian framework, where the structure is meshed by massive elements interacting through force potentials. The equations of motion are directly integrated, in our case using a 5$^{\mathrm{th}}$ order Gear predictor-corrector scheme, with time increments between 10$^{-6}$s and 10$^{-5}$s (see Masoero, Wittel et al., 2010). DEM is an equivalent formulation to Finite Elements, converging to the same numerical solution of the dynamics if identical force-displacement laws are implemented. A detailed description of the algorithm for 3D systems can be found in (Masoero, Wittel et al., 2010); \citep{Masoero-PHD-2010}, together with a discussion on the applicability. For this work, the code was restricted to 2D by allowing only two displacements and one rotation in the vertical plane. In (Masoero, Vallini et al., 2010), the \DEM~model is tested against dynamic energy-based collapse analyses of a continuous horizontal beam suddenly losing a support. In the appendix, we compare our \DEM~results to experimental observations of a 2D frame undergoing quasi-static column removal \citep{Yi_Kunnath_al_ACI-2008}. To the best of our knowlede, literature still lacks on experiments of dynamic collapse of framed structures due to accidental damage. 

In the following we will review only the essentials of our model, focusing on the details tha are relevant for the application to 2D frames. We assume simplified force-displacement laws for the beam element and for the Hertzian contacts. Predicting collapse of real structures would require more specialized interaction as compared to here, for example using the fiber approach for the cross sections. By contrast, we are interested in fundamental mechanisms of damage propagation within complex structural systems. In this research perspective, and according to a basic principle of Statistical Mechanics, minimizing the complexity of local interactions improves the interpretation of the systemic response. Despite the strong assumptions, in the Appendix we show that our model can match reasonably well with with experimental observations.
\subsection*{Structural representation}
\begin{figure}[htb]
\begin{center}
\includegraphics[scale=1]{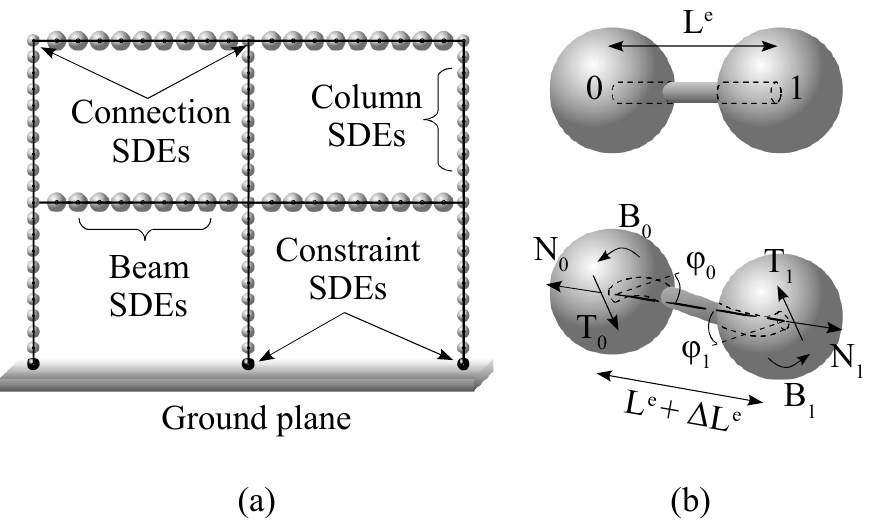}
\caption{a) Typical \DEM~mesh of a 2D frame. The centers of the spheres are connected by b) Euler-Bernoulli beam elements, shown in undeformed (above) and deformed state (below).}\label{Fig_2D_mesh}
\end{center}
\end{figure}

In a first step, the structure needs to be assembled by discrete elements and beams. \fignameSP\ref{Fig_2D_mesh}\subfig a shows the four types of Spherical Discrete Elements (\SDE) that we employed. Columns and beams are made of 9 \SDE s, respectively with diameter $\DiamCol=0.8\HeightCol$ and $\DiamBeam=0.8\HeightBeam$, slightly smaller than the distance between them to prevent contact form occurring before local rupture. Constrained \SDE s and connection \SDE s have same diameter $\DiamCol$ as column \SDE s. The constrained \SDE s are clamped to a plane that represents the ground by means of the Hertzian contact model, discussed further in this section.

\subsection*{Euler-Bernoulli beam elements}
Pairs of \SDE s are connected by \textit{Euler-Bernoulli beam elements - (\EBE)} that, when deformed, transmit forces and moments to their edge nodes, locally labeled 0 and 1 (see \fignameSP\ref{Fig_2D_mesh}\subfig b). The mass $M$ of an \SDE~is defined on the basis of the \EBE s connected to it. Namely $M = \sum\limits_{e} 1/2~\SpecWeiRC\EBEArea\EBELen$, where $e$ labels the generic \EBE~connected to \SDE, and $\EBEArea$ is the cross sectional area of the structural element corresponding to the \EBE. The external load $\ExtLoad$ is introduced adding a mass $\ExtLoad\EBELen/\Gravity$ to the beam SDEs. $\ExtLoad$ is not treated directly as a force to avoid downward accelerations of the \SDE s greater than gravity $\Gravity$ during free fall. 

For sufficiently small deformations, the \EBE s are linear elastic and exert a force $\AxForce$ proportional to the elongation $\AxStrain$ and directed along the $\overline{01}$ segment, a shear force $\Shear$ proportional to the sum of the nodal rotations $6\left(\Rot_0+\Rot_1\right)$, and a bending moment $\Bend$ proportional to the nodal effective rotations, defined as $\EffRot_0=4\Rot_0+2\Rot_1$, and $\EffRot_1=2\Rot_0+4\Rot_1$. Furthermore, we introduce damping by forces and moments directed opposite to $\AxForce$, $\Shear$, and $\Bend$, and proportional to the time derivative of $\AxStrain$ with factor $\EBEDampL=100$Ns/m, and of $\Rot_0$,$\Rot_1$ with factor $\EBEDampB=10$Nms. Geometric nonlinearity due to large displacements is considered by referring rotations and elongation to the $\overline{01}$ segment. In the small deformations regime of our simulations, $\AxForce$ is with good approximation equal to the axial force inside the \EBE, and thus perpendicular to $\Shear$. 

If $\AxForce$ overcomes a yield threshold in tension $\YieldNT$ or under compression $\YieldNC$, the ideally plastic regime is entered and plastic axial strain $\PlastAxStrain$ is applied to maintain $\AxForce=\YieldNT$ or $\AxForce=\YieldNC$. Neglecting the contributions of concrete in tension and of steel in compression, we set the yield thresholds in terms of $\AxForce$ and $\AxStrain$ to:
\begin{align}
 &\YieldNT= \AreaReinf \YieldStressSteel 	\;\;	\rightarrow 	
		\;\;	\EBEYieldEpsT= \frac{\YieldNT}{\EBEArea\YoungERC} =  \frac{\ReinfFrac \YieldStressSteel}{\YoungERC} 	\;\;,\; \mathrm{and}\\
 &\YieldNC= \left(\EBEArea - \AreaReinf \right) \CStrengthEC		\;\;	\rightarrow 	
		\;\;	\EBEYieldEpsC= \frac{\YieldNC}{\left(\EBEArea - \AreaReinf \right)\YoungERC} \approx  \frac{\CStrengthEC}{\YoungERC} 	\;\;.
\end{align}
Ideally plastic regime in bending is entered when $\left|\Bend\right|\ge\YieldBend$. We obtain the bending yield threshold $\YieldBend$ and the corresponding yielding effective rotation $\YieldEffRot$, neglecting the strength contribution of concrete and assuming a lever arm between upper and lower reinforcement equal to the height $\HeightCross$ of the cross section:
\begin{equation}\label{Eq_2D_By}
\YieldBend = \TensReinfFrac \ReinfFrac \EBEArea \YieldStressSteel \HeightCross +\ExtraYieldBend  \;\;\;				\mathrm{and} 	\;\;\;	\YieldEffRot= \frac{\YieldBend}{\YoungERC\EBEInertia} L^e	\;\;.	\\
\end{equation}
$\EBEInertia$ is the cross sectional moment of inertia of the \EBE, and $\TensReinfFrac$ is the fraction of reinforcement in tension (${3}/{8}$ for columns and ${1}/{2}$ for beams, as in \fignameSP\ref{Fig_Struct}\subfig b). $\ExtraYieldBend$ considers the beneficial compression effect compression in the \EBE. We set $\ExtraYieldBend$ assuming bending carried by the reinforcement alone, and that the strain $\StrainReinf\left(\ExtraYieldBend\right)$ in the reinforcement put under tension by $\ExtraYieldBend$ equals the compressive strain $\StrainReinf\left(\AxForce\right)$ due to $\AxForce<0$, namely:
\begin{equation}\label{Eq_2D_DBy}
\StrainReinf\left(\AxForce\right) =\StrainReinf\left(\ExtraYieldBend\right) 	\;\;\rightarrow\;\; 
	-\frac{\AxForce}{\EBEArea \YoungERC} = \frac{\ExtraYieldBend}{\TensReinfFrac \ReinfFrac \EBEArea \HeightCross \YoungESteel} \;\;.	 
\end{equation}
In this way, eventual tension $\AxForce>0$ inside the \EBE~produces negative $\ExtraYieldBend$, and thus reduces $\YieldBend$. When yielding in bending occurs, plastic rotations are added at the edge nodes of the \EBE. If only $\left|\Bend_i\right|$, with $i=0,1$, is greater than $\YieldBend$, then only $\RotPlast_i$ is applied to restore $\left|\Bend_i\right|=\YieldBend$. Differently, if both $\left|\Bend_0\right|$ and $\left|\Bend_1\right|$ are greater than $\YieldBend$, both $\RotPlast_0$ and $\RotPlast_1$ are applied to restore  $\left|\Bend_0\right|=\left|\Bend_1\right|=\YieldBend$. For the sake of simplicity, we assume yielding in bending uncoupled from yielding in axial direction. Furthermore, we neglect yielding due to shear because small plastic deformations are generally associated with shear. 

We consider an \EBE~failed when excessive $\PlastAxStrain$ and $\RotPlast$ are cumulated. For this purpose, the coupled breaking criterion:
\begin{align}
& \frac{\PlastAxStrain}{\left(\EloThresh-\EBEYieldEpsT\right)} 	+ \max_{i} \left( \frac {\left|\RotPlast_i\right|} {\RotThresh} \right)    \ge 1 \;\;\;\; \mbox{if} \;\;\PlastAxStrain>0  \;\;\mathrm{and}  \label{Tbreak}\\	
 -&\frac{\PlastAxStrain}{\left|\CompThresh-\EBEYieldEpsC\right|} + \max_{i} \left( \frac {\left|\RotPlast_i\right|} {\RotThresh} \right) \ge 1	\;\;\;\; \mbox{if} \;\;	\PlastAxStrain<0 \;\;,\label{Cbreak}
\end{align}
is employed. $\EloThresh$, $\CompThresh$, and $\RotThresh$ are the maximum allowed plastic elongation, shortening, and rotation in uncoupled conditions. We consider high plastic capacity of the structural elements setting $\EloThresh=2\UltStrainSteel$, $\CompThresh=2\UltStrainRC$, and $\RotThresh=0.2$rad (see \tabnameSP\ref{tabMecPar}). Failed \EBE s are instantly removed from the system. We neglect ruptures due to shear assuming that, in agreement with a basic principle of capacity design, a sufficient amount of bracings ensures the necessary shear strength. 
\subsection*{Inter-sphere contact}
The Hertzian contact model is employed for the \SDE s to consider collisions between structural elements. The model consist of repulsive forces between partially overlapping \SDE s, damped by additional forces proportional and opposite to the overlapping velocity. We also set tangential forces that simulate static and dynamic friction, as well as damping moments opposed to the relative rolling velocity. A similar Hertzian contact model is also employed for \SDE s colliding with the ground plane. In the following simulations we employ contact parameters that can be found in \citep{Masoero-PHD-2010}. We do not transcribe them because impacts do not affect significantly the collapse mechanisms sudied here. Nevertheless, in general simulation algorithms for progressive collapse should consider impacts, because initial damage located at upper stories generates falling debris, and because impacts can drive the transition from partial to total collapse (see (Masoero, Wittel et al., 2010) and \citep{Bazant_Zhou-2002}). In granular dynamics, the contact parameters are generally set referring to the material of the grains \citep{Poschel_Schwager-2005}. In our model the \SDE s represent large heterogeneous portions of structural elements, for which there are not conventionally defined contact parameters so far. We emply parameters yielding a qualitatively realistic dynamics (e.g.~the elements do not rebound or pass through each other), and chosen from sets of possible one that were defined through preliminary studies. Such studies also indicated that the collapse loads of a beam due to debris impact varies of less than 15\% upon orders of magnitude change in the contact parameters. 
\section*{Simulating progressive collapse}
The simulations are organized into two steps: first the structure is equilibrated under the effect of $\ExtLoad$ and gravity, then the \EBE s inside the damage area are suddenly removed, and the subsequent dynamic response is simulated. Our aim is to quantify three \textit{collapse loads}: 
\begin{itemize}
\item $\UltLoadInt$:  maximum static load  that the intact structure can carry;
\item $\CritLoad$: minimum \textit{critical load} that causes dynamic collapse after damage. Applied statically to the intact structure first, it is then kept constant during the post-damage dynamic response.
\item $\PartTotLoad$: minimum load corresponding to total collapse after damage. By definition, $\CritLoad \le \PartTotLoad \le \UltLoadInt$.
\end{itemize}
In our \DEM~model we do not have a straightforward unique measure of load, because the mass of the \SDE s depends on the external load $\ExtLoad$ and on the self weight of the structural elements. The mass of the beam \SDE s effectively acts as a distributed horizontal load. On the other hand, the columns at each story transmit vertical concentrated forces either to other columns at a lower story, or to the horizontal transfer beam over the damage area. Therefore we introduce a load measure that we call \textit{equivalent load} $\EqLoad$, applied to the massless structure and analytically related to the geometry, the mass, and the activated collapse mechanism of the frames in the simulations. Namely, $\EqLoad$ is defined to produce the same static effect as the various masses and concentrated forces of the simulation frames, at the critical points where collapse is triggered. The derivation of the analytical expressions used in this work is shown in \citep{Masoero-PHD-2010}.

For each analyzed structure, we first apply the entire structural mass. In a subsequent step, the external load $\ExtLoad$ is increased until the intact structure collapses in static conditions. The collapse mechanism indicates what equivalent load expression should be used to compute $\UltLoadInt$. Then we slightly decrease $\ExtLoad$, equilibrate, introduce the damage, and calculate whether dynamic progressive collapse is triggered and to what an extent. Performing several simulations with progressively smaller $\ExtLoad$, the final extent of collapse changes from total to partial, and we employ again an adequate equivalent load to compute $\PartTotLoad$. If the structure collapses even when $\ExtLoad$ is reduced to zero, we start reducing the specific weight of the structural elements, i.e.~the structural mass. When dynamic collapse does not occur anymore, an adequate equivalent load provides $\CritLoad$. Once we obtain the collapse loads, we estimate the progressive collapse resistance referring to the \textit{residual strength fraction} $\ROne=\CritLoad/\UltLoadInt$. Actually, progressive collapse resistance is more directly related to $\CritLoad$, but the advantage of $\ROne$ is that it can not be improved by simply strengthening the structural elements, which would increase both $\CritLoad$ and $\UltLoadInt$. Robustness-oriented structural optimization is required to increase $\ROne$, which therefore is a good indicator to compare different structural solutions.
\subsection*{Bending collapse}
\begin{figure}[htb]
\begin{center}
\includegraphics[scale=1]{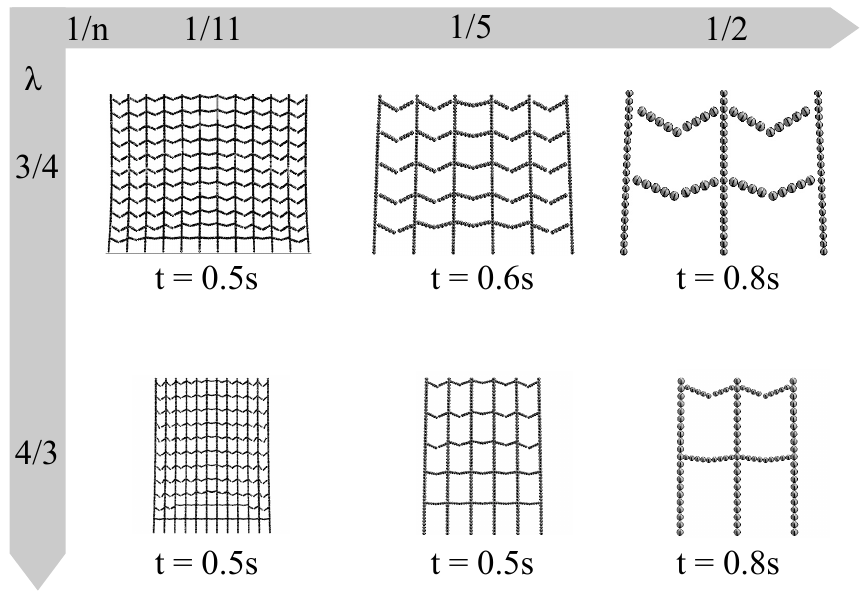}
\caption{Static bending collapse mechanism before damage. Time $t=0$s corresponds to the first breaking of an EBE.}\label{Fig_2D_BendColl_before}
\end{center}
\end{figure}

\begin{figure}[htb]
\begin{center}
\includegraphics[scale=1]{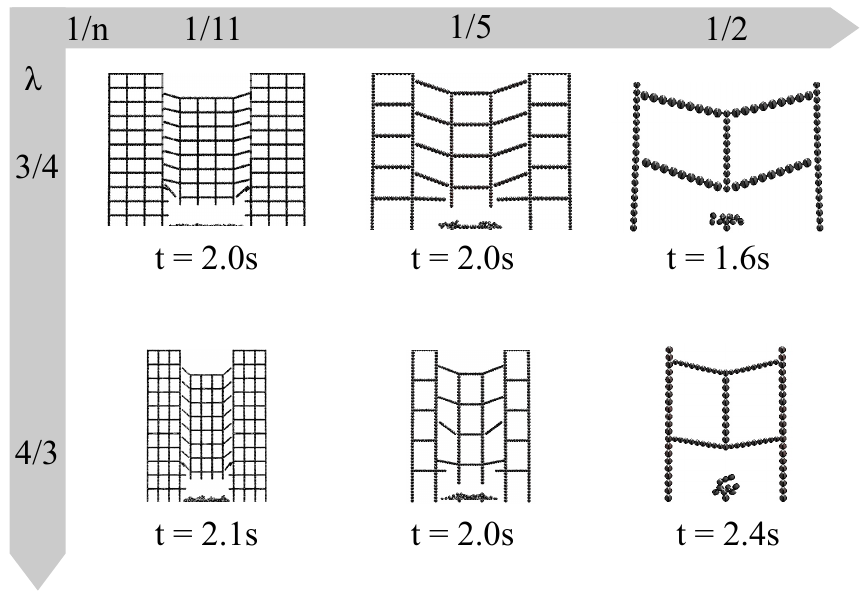}
\caption{Dynamic bending collapse mechanism after damage. Time $t=0$s corresponds to the application of the initial damage.}\label{Fig_2D_BendColl}
\end{center}
\end{figure}

In our model, the bending yield threshold $\YieldBend$ does not depend on the strength of concrete $\CStrengthEC$. Therefore, setting the high value $\CStrengthEC= 35$N/mm$^2$, the mainly compressed columns get much stronger than the horizontal beams, that fail in bending (see \fignamesSP\ref{Fig_2D_BendColl_before},\ref{Fig_2D_BendColl}). The resulting collapse mechanisms resemble triple-hinge and four hinges mechanisms, reflecting the large plastic capacity of the structural elements.

\begin{figure}[htb]
\begin{center}
\includegraphics[scale=1]{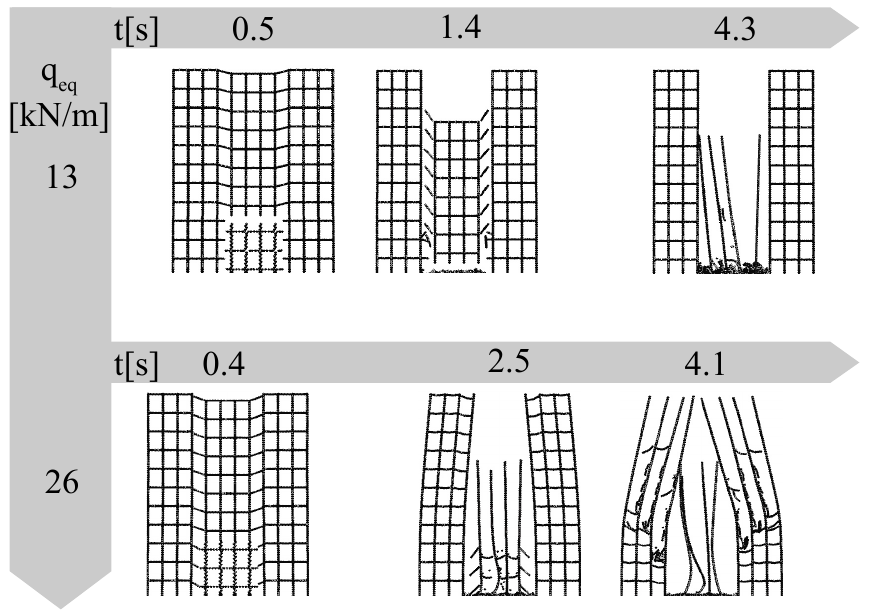}
\caption{Snapshots of partial ($q_{eq}$=13kN/m) and total ($q_{eq}$=26kN/m) bending collapse after damage of a frame with very strong columns, $\NCell$=11, and $\SlendCell=1.33$. Time $t=0$s corresponds to the application of the initial damage.}\label{Fig_par-tot-bend}
\end{center}
\end{figure}

If the initial damage triggers a bending mechanism, frames with $\NCell=2$ undergo total collapse, while frames with lower hierarchical level $1/\NCell$ initially suffer only a local collapse  (see \fignameSP\ref{Fig_par-tot-bend}). The local collapse can nevertheless evolve to total collapse, if high applied load and plastic capacity cause the falling central part of the structure to dynamically drag down the lateral portions (Masoero, Wittel et al., 2010).

\begin{figure}[htb]
\begin{center}
\includegraphics[scale=1]{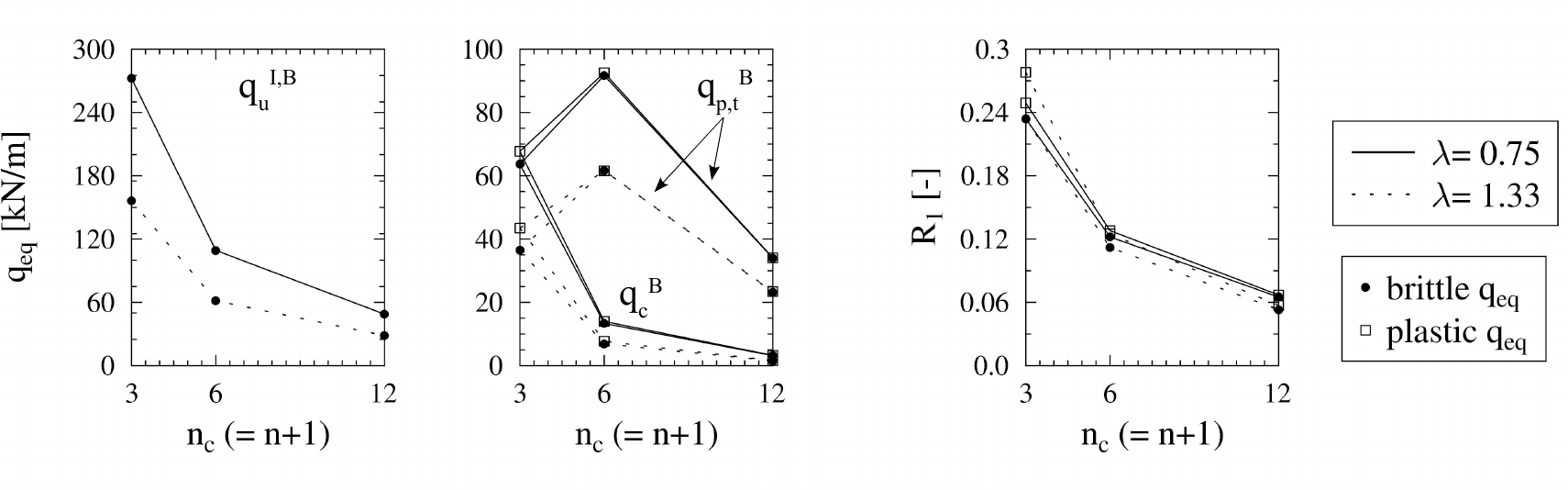}
\caption{Equivalent collapse loads and residual strength fraction $\ROne$ for frames that undergo bending progressive collapse.}\label{Fig_mu-RSR-bend}
\end{center}
\end{figure}

\begin{figure}[htb]
\begin{center}
\includegraphics[scale=1]{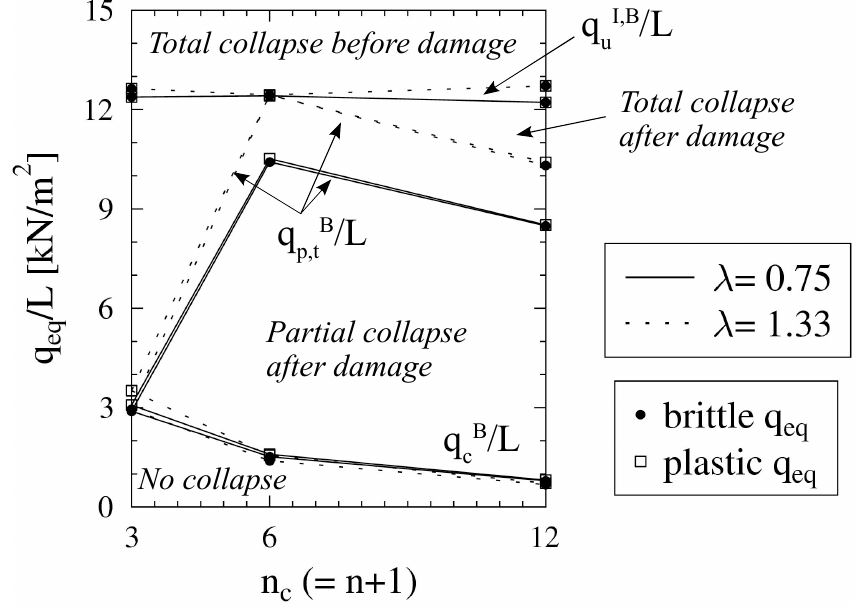}
\caption{Collapse loads in \fignameSP\ref{Fig_mu-RSR-bend} divided by $\Bay$, considering the 2D frames as part of regular 3D structures.}\label{Fig_divL_mu-bend}
\end{center}
\end{figure}

The collapse loads, expressed in terms of equivalent loads $\EqLoad$, are summarized in \fignameSP\ref{Fig_mu-RSR-bend} as a function of the hierarchical level $1/\NCell$, for different slenderness of the structural cells $\SlendCell$. In \fignameSP\ref{Fig_mu-RSR-bend}, superscript $B$ indicates bending collapse mechanism. We employ equivalent loads referring to  perfectly brittle or perfectly plastic bending failure (see the Appendix). The collapse loads decrease with $\SlendCell$, i.e.~a slender structure seems weaker, and increase with $1/\NCell$, i.e.~hierarchical frames are stronger. The residual strength fraction $\ROne$ does not depend on $\SlendCell$, while hierarchical structures with low $\NCell$ are more robust than homogeneous ones (see \fignameSP\ref{Fig_mu-RSR-bend}). In fact, the concentration of bending moment at the connection between a beam hanging above the damage area and the first intact column depends on the \textit{number} of removed columns. In the simulations, we remove a constant fraction of one third of the columns on a horizontal line (see \fignameSP\ref{Fig_Struct}). Therefore homogeneous structures lose more columns and are less robust toward the bending collapse mechanisms. On the other hand, since the number of removed columns is decisive, we expect that the hierarchical level does not influence $\ROne$ toward bending collapse in case of single column removal.
Finally we consider the 2D frame as part of a regular 3D structure and divide the collapse loads in \fignameSP\ref{Fig_mu-RSR-bend} by $\Bay$, i.e.by the tributary length of the beams in the direction perpendicular to the frame. In this way, collapse loads per unit area are obtained in \fignameSP\ref{Fig_divL_mu-bend}, showing that: $\SlendCell$ does not influence $\CritLoad/\Bay$ and $\UltLoadInt/\Bay$; structures with slender cells are less likely to collapse entirely; $\UltLoadInt/\Bay$ is independent from the hierarchical level; $\CritLoad/\Bay$ is proportional to $1/\NCell$.

\subsection*{Pancake collapse}
\begin{figure}[htb]
\begin{center}
\includegraphics[scale=1]{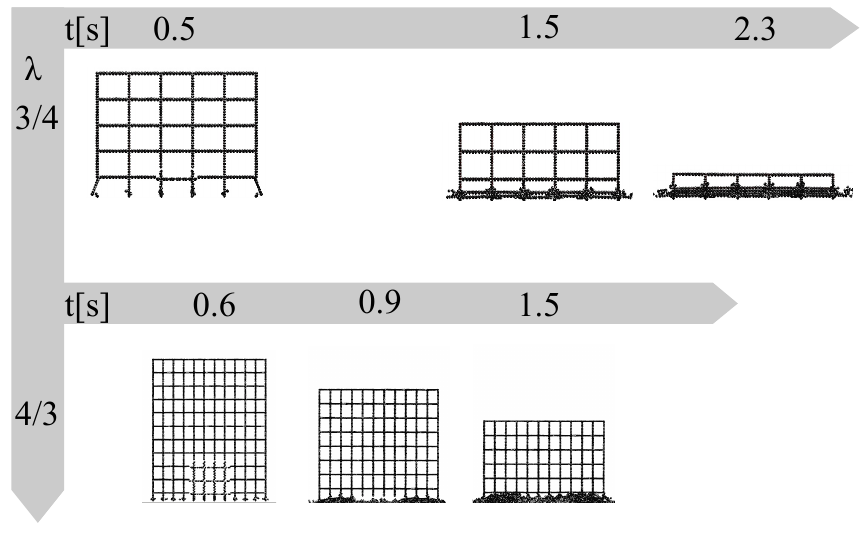}
\caption{Snapshots of pancake collapse after damage for frames with $\NCell$=5, $\SlendCell=0.75$, and $\NCell$=11, $\SlendCell=1.33$.}\label{Fig_2D_pancake}
\end{center}
\end{figure}

Progressive compressive failure of the columns, also called pancake collapse, occurs when we set the compressive strength of concrete to a small value $\CStrengthEC=0.35$N/mm$^2$ (see \fignameSP\ref{Fig_2D_pancake}). This choice is unphysical but allows us to separate the effect of strength reduction from that of stiffness reduction in the columns. More realistic scenarios would involve columns with small cross section and highly reinforced, tall beams. 

The columns immediately next to the damage area are the first to fail under compression, and then progressive collapse spreads horizontally to the outside. We employ equivalent loads $\EqLoad$ referring to the two limit cases of \textit{local} and of \textit{global} pancake collapse. Local pancake collapse occurs when the bending stiffness of the beams is very low and when the compressive failure of the columns is very brittle. In this case, the overload after damage is entirely directed to the intact columns that are closer to the damage area, and collapse propagates by \textit{nearest neighbor} interactions. On the other hand, high stiffness of the beams and large plastic capacity of the columns induce \textit{democratic} redistribution of overload between the columns. Consequently, the columns crush simultaneously triggering global pancake collapse. The collapse dynamics recorded in our simulations resembles global pancake. Note that in the studied framed structures, all the columns have identical compressive strength without disorder. Therefore, once the first two columns crush, pancake collapse can not be arrested. Nevertheless, at some $\CStrengthEC>0.35$N/mm$^2$ our frames undergo partial collapse because the progressive failure of the columns can be arrested by the initiation of bending collapse.

\begin{figure}[htb]
\begin{center}
\includegraphics[scale=1]{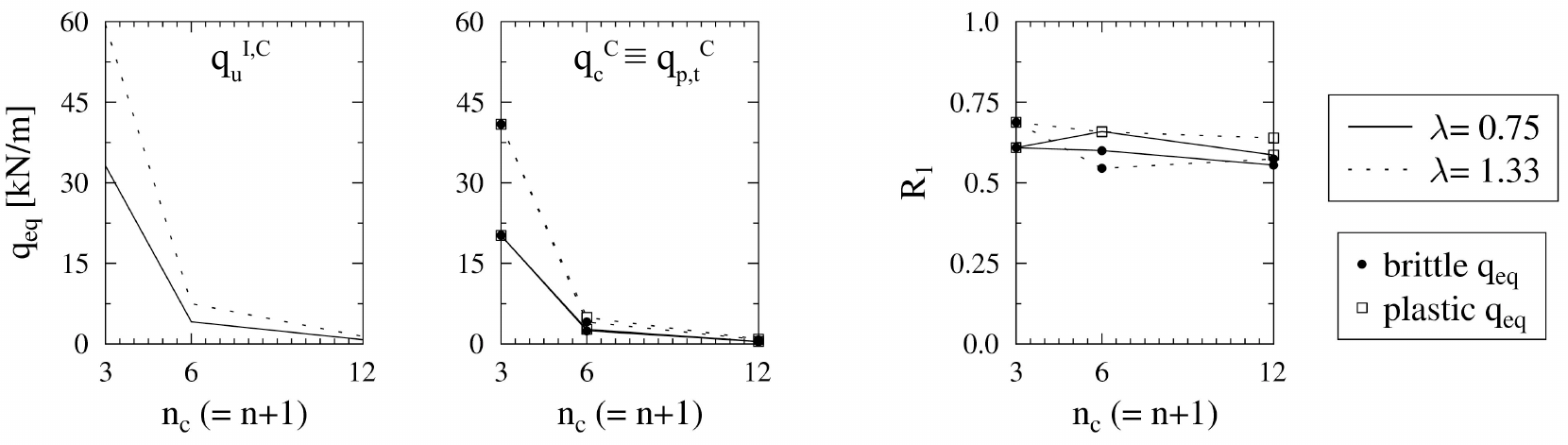}
\caption{ Equivalent loads and residual strength fraction for the studied structures undergoing pancake collapse.}\label{Fig_2D_mu-RSR_pank}
\end{center}
\end{figure}

\fignameSP\ref{Fig_2D_mu-RSR_pank}, where superscript $C$ indicates pancake collapse, shows that the collapse loads increase with the structural slenderness $\SlendCell={\Height}/{\Bay}$, because the columns have tributary area related to $\Bay^2$ and compressive strength proportional to $\Height^2$. Furthermore, hierarchical structures with small $\NCell$ appear to be stronger than homogeneous ones both in terms of $\UltLoadInt$ and of $\CritLoad$. Finally, the residual strength fraction $\ROne$ toward pancake collapse is remarkably higher than that toward bending collapse (cf.~\fignameSP\ref{Fig_mu-RSR-bend}), and is neither influenced by the hierarchical level $1/\NCell$, nor by $\SlendCell$. In fact, $\ROne$ toward global pancake mode is related to the \textit{fraction} of columns that are initially removed at one story. In our simulations, we always remove one third of the columns at one story, and obtain the constant value $\ROne\approx0.6$, slightly smaller than a theoretical 2/3 because of dynamics.

\section*{Conclusions}
We showed how the dynamic strength after damage $\CritLoad$ of 2D frames depends on the activated collapse mechanism. We can now drive a series of conclusions regarding the effect of damage extent, structural slenderness, and topological and mechanical hierarchy. Bending collapse provokes a local intensification of bending moments at the connections between the transfer beams above the damage area and the first intact column. Consequently, $\CritLoad$ and the residual strength fraction $\ROne$ decrease with the number of removed columns. In analogy with fracture mechanics, structures that are prone to bending collapse correspond to notch sensitive materials, and the number of removed columns corresponds to the crack width \citep{Chiaia_Masoero-2008}. If global pancake collapse is triggered, $\CritLoad$ and $\ROne$ decrease with the fraction of removed columns, which is analogous to plastic failure of materials that are not notch sensitive. Consistently, $\ROne$ corresponding to global pancake collapse is remarkably larger than that corresponding to bending collapse.

The structural slenderness $\lambda=H/L$ affects in general the collapse loads for both bending and pancake collapse modes. The effect of $\lambda$ depends on the scaling of cross section and reinforcement of the structural elements, with the beam length $L$ and with the column height $H$ (see the analytical results in \citep{Masoero-PHD-2010,Masoero_Chiaia_submitt-2011}, regarding the simulations in this paper). Nevertheless $R_1$ turns out to be independent from $\lambda$, because $R_1$ is the ratio between two collapse loads with same scaling respect to $L$ and $H$.

Considering structural topology, in case of bending collapse hierarchical structures are more robust toward initial damage with fixed spatial extent (e.g.~explosion, impact), and as robust as homogeneous structures toward single column removal (e.g.~design error). The reason is that $\ROne$ toward bending collapse decreases with the number of removed columns at one story. This confirms the analogy with fracture mechanics, where notch sensitive hierarchical materials are tougher than homogeneous ones \citep{Lakes-1993}. On the other hand, considering global pancake collapse, structural hierarchy does not influence $\ROne$ toward initial damage with fixed spatial extent, while hierarchical structures are more sensitive than homogeneous ones to single column removals. This is due to $\ROne$ toward global pancake collapse decreasing with the fraction of removed columns. \fignameSP\ref{Fig_2D_mu-RSR_pank} shows that damaged frames undergoing global pancake collapse can carry the $\ROne\approx60\%$ of the static ultimate load of the intact structure $\UltLoadInt$. Since well designed structures can carry a $\UltLoadInt$ remarkably greater than the environmental load expected when an accidental event occurs, $\ROne$ related to global pancake collapse can ensure structural robustness for most of the practical cases \citep{Masoero-PHD-2010}. On the other hand, $\ROne$ related to bending collapse is generally much smaller, making structures vulnerable to accidental damage. 

In this work we considered idealized structures, with simplified geometry and mechanical behavior of the elements. Reducing local complexity enables a better interpretation of the coral system response to damage. This study provides a basis of knowledge preceding the incorporation of more details and degrees of freedom, to investigate further aspects of progressive collapse. Shear failures can cause brittle ruptures and reduce the collapse resistance of large structural elements. Different locations of the initial damage may activate different collapse mechanisms. For example, damaging the upper stories would cause debris impacts, while removing external columns reduces $q_c$ without producing significant lateral toppling \citep{Calvi_Master_thesis-2010}. The DEM algorithm was already applied to 3D structures in (Masoero, Wittel et al., 2010), showing that the bending and pancake collapse mechanisms persist also in 3D. On the other hand, in 3D structures the horizontal floor slabs improve the horizontal redistribution of loads and the catenary action, increasing the strength toward bending collapse and impacting debris (see the Appendix and, e.g.~, \citep{Vlassis_Izzudin-2008}). It is worth noting that horizontal ties and diaphragms increase the strength both after and before damage, causing a compensation that limits the effect on $R_1$. Finally, future works can incorporate a detailed description of structural connections, which are crucial for energy dissipation, catenary effect, and compartmentalization.

Coming back to the central theme of structural hierarchy, our results already suggest that hierarchical structures are more robust toward accidental damage. An optimal solution would be to design: 1) a primary frame made of few large elements, with columns weaker than the beams, and 2) a secondary structure, made of many smaller elements, which defines the living space and follows traditional design rules. The primary frame would provide topological hierarchy, maximizing $\ROne$ toward bending collapse and enabling new possible compartmentalization strategies. The strong beams and weak columns of the primary frame would favor pancake collapse over bending collapse, and improve the vertical compartmentalization of high-rise buildings against falling debris. On the other hand, in real structures, the beams generally fail before the columns, and imposing the opposite is expensive. Nevertheless, designing a strong-beam weak-column behavior \textit{only for the primary frame} can significantly limit the extra cost. Hierarchical structures can be a novel and somehow counterintuitive feature of robustness-oriented capacity design. Planning structural hierarchy requires understanding the complex system response to local damage, and should drive the design process since the very beginning. By contrast, traditional design is focused on local resistance against ordinary actions, and considers robustness toward accidents only at the end. This generally leads to non-hierarchical structures with strong columns and poorly understood system behavior. In addition, anti-seismic capacity design requires plastic failure of the beams to precede columns rupture (see e.g.~\citep{Byfield-2004}). Overcoming these contradictions is a challenge toward optimizing structures against exceptional events.

\pagebreak
\newpage
\appendix
\section*{Appendix: experimental benchmark}
In this appendix, we compare the numerical predictions of our DEM model with the experimental observations in \citep{Yi_Kunnath_al_ACI-2008}. We also briefly discuss some effects of catenary actions on collapse resistance.

\begin{figure}[htb]
\begin{center}
\includegraphics[scale=1]{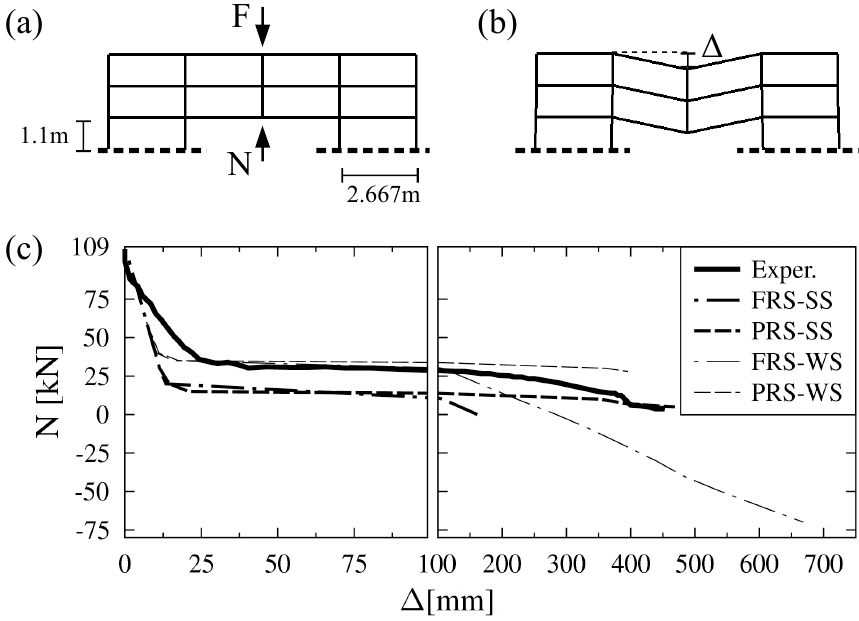}
\caption{(a) Geometry of the model frame and (b) generic bending collapse mechanism. (c) Experimental reaction curve, and predictions from our model. All the predicted curves are plotted until a first EBE fails in tension, with exception for the FRS-SS curve, truncated earlier due to evident divergence from the experimental result. If fully reactive cross sections FRS are assumed, collapse occurs when $N<<0$, i.e.~$N$ contributes to pulling down the structure instead of contrasting $F$.}\label{Fig_exper}
\end{center}
\end{figure}

\paragraph{Experiments}
The experimental setup in \citep{Yi_Kunnath_al_ACI-2008} consists of a plane frame made of reinforced concrete (see \fignameSP\ref{Fig_exper}(a)). Columns are square in section (200x200mm), beams are rectangular (200mm tall, 100mm wide). Everywhere, the longitudinal reinforcement is symmetrically distributed within the cross section (4$\phi$12 steel bars). The strength and ultimate strain of concrete and steel are specified in \citep{Yi_Kunnath_al_ACI-2008}, while the elastic moduli are not. The mid column at the first floor is replaced by jacks that provides an upward vertical force $N$. In the middle of the top floor, a servo-hydraulic actuator applies a constant downward vertical force $F$=109kN, to represent the self weight of upper stories. Initially $N=F=$109kN, and then it is progressively reduced to reproduce quasi-static column loss, until a bending mechanism triggers collapse (see \fignameSP\ref{Fig_exper}(b)). During the experiments, the increasing values of the midspan inflection $\Delta$ is plotted against $N$, to get the force-displacement reaction curve. The integral of the curve represents the energy dissipation capacity, which relates to the dynamic strength of the structure with respect to the activated collapse mechanism. 

\paragraph{Model description}
Our target is to capture the experimental reaction curve $N-\Delta$ through DEM simulations. The parametrization of our model, based on the geometry and mechanical data in \citep{Yi_Kunnath_al_ACI-2008}, is straightforward. Therefore, we focus on the discrepancies between model and experimental inputs, and a few necessary additional assumptions. Regarding the overall geometry, we consider all the columns to be equally tall (1,100mm), while in the experiments the columns at the first floor were taller (1,567mm). This discrepancy should not have a significant effect on the collapse mechanism and the strength. The mechanical behavior of the real steel bars was strain hardening, with yielding at 416MPa, and rupture at 526MPa. In our model, we consider two limit cases of elastic-perfectly plastic behavior of the steel bars: \textit{weak steel} ``WS'' with yielding threshold set at 416MPa, and \textit{strong steel} ``SS'' yielding at 526MPa. \citep{Yi_Kunnath_al_ACI-2008} provide two measures of the ultimate tensile strain $\delta$ of the steel bars. We employ $\delta_{10} = 23\%$, which  was measured on a longer bar segment, because in our simulation the strain develops within relatively long Euler-Bernoulli Beam Elements, EBEs. We assume Young moduli $E_s=200$GPa for the steel, and $E_c=30$GPa for the concrete. In order to better understand the development of catenary actions, we consider two limit cases of cross section behavior under tension: \textit{fully reacting sections} ``FRS'', where the concrete always contributes to the tensile stiffness, and \textit{partially reacting sections} ``PRS'' , where the concrete cracks and only the steel provides axial stiffness as soon as the cross section goes in tension. Furthermore, in order to focus exclusively on ruptures due to tensile strain in the steel, we allow for an infinite rotation capacity of the cross sections.

\paragraph{Simulations and results}
We subject our model frames to gravity, but remain in the quasi-static regime by adding a high viscous damping force proportional to the velocity of each Spherical Discrete Element. We repeat numerous simulations with fixed $F=109$kN and $N$, ranging from $N=109$kN to values that are small enough to cause the quasi-static rupture of at least one EBE. We track the midspan deflection $\Delta\left(N\right)$for comparison in \fignameSP\ref{Fig_exper}(c).

\paragraph{Discussion}
In the experimental results, as $N$ decreases, the system crosses several stages: (I) linear elastic $\Delta\lessapprox 5$mm, (II) elasto-plastic $\Delta< 22$mm, (III) plastic hinges $\Delta< 140$mm, (IV) catenary action $\Delta< 450$mm, and (V) collapse. The transition from elastic to elasto-plastic is not evident from the curve, as well as that from plastic hinge to catenary action. By contrast, plastic hinges formation is clearly marked by a sudden change of slope at $\Delta \approx 40$mm. Our simulations do not capture the initial elasto-plastic stage because we do not model the non-linear elasto-plastic behavior of concrete. This leads to an overestimation of the stiffness $dN/d\Delta$ before the formation of the plastic hinges. Nevertheless, the additional strain energy produced by this approximation is negligible when compared to the energy dissipated in the subsequent stages, ie.~the overestimation of the initial stiffness is irrelevant for the actual dynamic collapse. Assuming weak steel WS, yielding at 416MPa, provides a good agreement with the experiment in terms of transition point to the plastic hinges stage. Considering fractured concrete under tension yields the PRS-WS curve, which underestimates the structural strength at large $\Delta$. The reason for this divergence can be that the steel hardens under strain, with reaction stress increasing from from 416MPa (WS) to 526MPa (SS). This interpretation is supported by the fact that the PRS-WS and PRS-SS curves envelop the experimental one. In particular, the PRS-SS curve reproduces well the last part of the experimental curve, as well as the collapse point. 

\paragraph{Conclusions} Despite strong simplifying assumptions in the formulation, our DEM model provides reasonably good quantitative predictions of the experimental results. For the simulations in the body of this paper, we always considered fully reactive cross sections with concrete that does not crack under tension. The FRS-WS curve in \fignameSP\ref{Fig_exper}(c) shows that this assumption leads to an overestimation of the static collapse strength against column removal ($\approx +70$\%). Let us conjecture that FRS induce the same strength increase of +70\% in the structure without column removal, i.e.~in $q_u^I$. From a heuristic application of energy conservation, one can estimate the dynamic collapse load after sudden column removal by considering the mean $N$ in the catenary stage: $N=20$kN from the experiment, and $N=-20$kN from the simulation with FRS-WS. Consequently, the increase in post-damage dynamic collapse strength $q_c$ due to FRS is approximately $\left(1-130/90\right)\cdot 100 = +44$\%. In conclusion, assuming fully reactive cross sections causes an \textit{underestimation} of the residual strength fraction $R_1 = q_c / q_u^I$, which our example quantifies as $\left(44/70-1\right) \approx -37$\%. However this assumption does not affect the main statement of this work on hierarchical structures.


\begin{thebibliography}{22}
\providecommand{\natexlab}[1]{#1}
\providecommand{\url}[1]{\texttt{#1}}
\expandafter\ifx\csname urlstyle\endcsname\relax
  \providecommand{\doi}[1]{doi: #1}\else
  \providecommand{\doi}{doi: \begingroup \urlstyle{rm}\Url}\fi

\bibitem[Alexander(2004)]{Alexander-2004}
S.~Alexander.
\newblock New approach to disproportionate collapse.
\newblock \emph{Struct. Eng.}, 82\penalty0 (23/24):\penalty0 14--18, 2004.

\bibitem[Ba\v{z}ant and Zhou(2002)]{Bazant_Zhou-2002}
Z.P. Ba\v{z}ant and Y.~Zhou.
\newblock Why did the {World Trade Center} collapse? - {S}imple analysis.
\newblock \emph{J. Eng. Mech.-ASCE}, 128\penalty0 (1):\penalty0 2--6, 2002.

\bibitem[{BS Eurocode 1}(2004)]{Eurocode_1-1-7}
{BS Eurocode 1}.
\newblock Actions on structures - part 1-7: General actions - accidental
  actions.
\newblock Technical Report EN 1991-1-7, BSI, 2004.

\bibitem[{BS Eurocode 8}(2004)]{Eurocode_8}
{BS Eurocode 8}.
\newblock Design of structures for earthquake resistance.
\newblock Technical report, BSI, 2004.

\bibitem[Byfield(2004)]{Byfield-2004}
M.P. Byfield.
\newblock Design of steel framed buildings at risk from terrorist attack.
\newblock \emph{Struct. Eng.}, 82\penalty0 (22):\penalty0 31--38, 2004.

\bibitem[Calvi(2010)]{Calvi_Master_thesis-2010}
A.~Calvi.
\newblock Il crollo delle torri gemelle: analisi dell'evento e insegnamenti
  strutturali.
\newblock Master's thesis, Politecnico di Torino, 2010.
\newblock (In Italian).

\bibitem[Carmona et~al.(2008)Carmona, Wittel, Kun, and
  Herrmann]{Carmona_Wittel-2008}
{H.A.} Carmona, {F.K.} Wittel, F.~Kun, and {H.J.} Herrmann.
\newblock Fragmentation processes in impact of spheres.
\newblock \emph{Phys. Rev. E}, 77\penalty0 (5):\penalty0 243--253, 2008.

\bibitem[Cherepanov and Esparragoza(2007)]{Cherepanov_Esparragoza-2007}
G.P. Cherepanov and I.E. Esparragoza.
\newblock Progressive collapse of towers: the resistance effect.
\newblock \emph{Int. J. Fract.}, 143:\penalty0 203--206, 2007.

\bibitem[Chiaia and Masoero(2008)]{Chiaia_Masoero-2008}
B.M. Chiaia and E.~Masoero.
\newblock Analogies between progressive collapse of structures and fracture of
  materials.
\newblock \emph{Int. J. Fract.}, 154\penalty0 (1-2):\penalty0 177--193, 2008.

\bibitem[DoD(2005)]{DoD_UFC-2005}
DoD.
\newblock {Unified Facilities Criteria (UFC): Design of Buildings to Resist
  Progressive Collapse}.
\newblock Technical report, Department of Defence, 2005.

\bibitem[GSA(2003)]{GSA-2003}
GSA.
\newblock {General Services Administration. Progressive Collapse Analysis and
  Design Guidelines for New Federal Office Buildings and Major Modernization
  Projects}.
\newblock Technical report, GSA, 2003.

\bibitem[Gulvanessian and Vrouwenvelder(2006)]{Gulvanessian-2006}
H.~Gulvanessian and T.~Vrouwenvelder.
\newblock Robustness and the eurocodes.
\newblock \emph{Struct. Eng. Int.}, 2:\penalty0 161--171, 2006.

\bibitem[Hamburger and Whittaker(2004)]{Hamburger_Whittaker-2004}
R.~Hamburger and A.~Whittaker.
\newblock Design of steel structures for blast-related progressive collapse
  resistance.
\newblock \emph{Modern Steel Constr.}, March:\penalty0 45--51, 2004.

\bibitem[Lakes(1993)]{Lakes-1993}
R.~Lakes.
\newblock Materials with structural hierarchy.
\newblock \emph{Nature}, 361:\penalty0 511--515, 1993.

\bibitem[Masoero(2010)]{Masoero-PHD-2010}
E.~Masoero.
\newblock \emph{Progressive collapse and robustness of framed structures}.
\newblock PhD thesis, Politecnico di Torino, Italy, 2010.

\bibitem[Masoero et~al.(2013)Masoero, Dar\`{o}, and
  Chiaia]{Masoero_Chiaia_submitt-2011}
E.~Masoero, P.~Dar\`{o}, and B.M. Chiaia.
\newblock Progressive collapse of 2d framed structures: an analytical model.
\newblock \emph{Engn. Struct.}, 54:\penalty0 94--102, 2013.

\bibitem[Pearson and Delatte(2005)]{Pearson_Delatte-2005}
C.~Pearson and N.~Delatte.
\newblock Ronan point apartment tower collapse and its effect on building
  codes.
\newblock \emph{J. Perf. Constr. Fac. -ASCE}, 19\penalty0 (2):\penalty0
  172--177, 2005.

\bibitem[P\"{o}schel and Schwager(2005)]{Poschel_Schwager-2005}
T.~P\"{o}schel and T.~Schwager.
\newblock \emph{Computational Granular Dynamics}.
\newblock Springer-Verlag GmbH, Berlin, 2005.

\bibitem[Starossek(2006)]{Starossek-2006}
U.~Starossek.
\newblock Progressive collapse of structures: nomenclature and procedures.
\newblock \emph{Struct. Eng. Int.}, 16\penalty0 (2):\penalty0 113--117, 2006.

\bibitem[Val and Val(2006)]{Val_Val-2006}
D.V. Val and E.G. Val.
\newblock Robustness of framed structures.
\newblock \emph{Struct. Eng. Int.}, 16\penalty0 (2):\penalty0 108--112, 2006.

\bibitem[Vlassis et~al.(2008)Vlassis, Izzuddin, Elghazouli, and
  Nethercot]{Vlassis_Izzudin-2008}
A.G. Vlassis, B.A Izzuddin, A.Y. Elghazouli, and D.A Nethercot.
\newblock Progressive collapse of multi-storey buildings due to sudden column
  loss-{Part II}: application.
\newblock \emph{Eng. Struct.}, 30\penalty0 (5):\penalty0 1424--1438, 2008.

\bibitem[Yi et~al.(2008)Yi, He, Xiao, and Kunnath]{Yi_Kunnath_al_ACI-2008}
W.-J. Yi, Q.-F. He, Y.~Xiao, and S.K. Kunnath.
\newblock Experimental study on progressive collapse-resistant behavior of
  reinforced concrete frame structures.
\newblock \emph{ACI Structural Journal}, 105\penalty0 (4):\penalty0 433--439,
  2008.

\end{thebibliography}
\end{document}